\documentclass[journal=nalefd,manuscript=letter]{achemso}
\usepackage{chemformula} 
\usepackage[T1]{fontenc} 
\usepackage[utf8]{inputenc}
\usepackage{graphicx}
\usepackage{dcolumn}
\usepackage{bm}
\usepackage{siunitx}
\usepackage{color}
\usepackage[normalem]{ulem}

\newcommand{\vect}[1]{\bm{#1}}

\author{Sara Pourjamal}
\affiliation{Department of Applied Physics, Aalto University School of Science, FI-00076 Aalto, Finland}

\author{Tommi K. Hakala}
\email{tommi.hakala@uef.fi}
\affiliation{Department of Applied Physics, Aalto University School of Science, FI-00076 Aalto, Finland}
\alsoaffiliation{Institute of Photonics, University of Eastern Finland, FI-80101 Joensuu, Finland}

\author{Marek Nečada}
\affiliation{Department of Applied Physics, Aalto University School of Science, FI-00076 Aalto, Finland}

\author{Francisco Freire-Fern\'{a}ndez}
\affiliation{Department of Applied Physics, Aalto University School of Science, FI-00076 Aalto, Finland}

\author{Mikko Kataja}
\affiliation{Department of Applied Physics, Aalto University School of Science, FI-00076 Aalto, Finland}
\alsoaffiliation {Institut de Ci\`{e}ncia de Materials de Barcelona (ICMAB-CSIC), Campus de la UAB, Bellaterra, Catalonia, Spain}

\author{Heikki Rekola}
\affiliation{Smart Photonic Materials, Faculty of Engineering and Natural Sciences, Tampere University, P.O. Box 541, FI-33101 Tampere, Finland}

\author{Jani-Petri Martikainen}
\affiliation{Department of Applied Physics, Aalto University School of Science, FI-00076 Aalto, Finland}

\author{P\"{a}ivi T\"{o}rm\"{a}}
\affiliation{Department of Applied Physics, Aalto University School of Science, FI-00076 Aalto, Finland}

\author{Sebastiaan van Dijken}
\email{sebastiaan.van.dijken@aalto.fi}
\affiliation{Department of Applied Physics, Aalto University School of Science, FI-00076 Aalto, Finland}

\title{Lasing in Ferromagnetic Plasmonic Arrays}

\begin{document}

\begin{abstract}
We report on lasing at visible wavelengths in arrays of ferromagnetic Ni nanodisks overlaid with an organic gain medium. We demonstrate that by placing an organic gain material within the mode volume of the plasmonic nanoparticles both the radiative and, in particular, the high ohmic losses of Ni nanodisk resonances can be compensated. Under increasing pump fluence, the systems exhibit a transition from lattice-modified spontaneous emission to lasing, the latter being characterized by highly directional and sub-nanometer linewidth emission. By breaking the symmetry of the array, we observe tunable multimode lasing at two wavelengths corresponding to the particle periodicity along the two principal directions of the lattice. Our results pave the way for loss-compensated magnetoplasmonic devices and topological photonics.
\end{abstract}

Plasmonic resonators and cavities provide small mode volumes and ultrafast light-matter interactions at the nanoscale. Interactions between emitters and plasmonic modes have been studied in both weak and strong coupling regimes\cite{TOR-15,BOZ-17}. Theoretical and experimental investigations on lasing in plasmonic systems have demonstrated the feasibility of compensating losses typical for metallic nanostructures\cite{bergman_surface_2003,hill_lasing_2007,stockman_spasers_2008,zheludev_lasing_2008,noginov_demonstration_2009,oulton_plasmon_2009,wuestner_overcoming_2010,ma_room-temperature_2011,khajavikhan_thresholdless_2012,lu_plasmonic_2012,zhou_lasing_2013,van_beijnum_surface_2013,meng_wavelength-tunable_2013,schokker_statistics_2015,dridi_lasing_2015,hakala_lasing_2017,RAM-17,WAN-18} and providing ultrafast operation speeds\cite{oulton_plasmon_2009,DAS-18}. Periodic arrays of metallic nanoparticles support collective surface lattice resonances (SLRs) that originate from radiative coupling of lossy single particle plasmon resonances with low-loss diffracted orders (DOs) of the lattice\cite{zou_silver_2005,KRA-08,auguie_collective_2008,humphrey_plasmonic_2014,WANG-18,KRA-18}. When the particle periodicity $p$ equals the wavelength of the radiation in the medium ($p=\lambda/n$, with $\lambda$ the wavelength in free space and $n$ the refractive index of the surrounding medium), the radiation fields at each particle interfere constructively, creating increased electric fields and phase correlations over several unit cells. Despite the plasmonic component, SLRs in arrays of noble metal nanodisks have particularly narrow linewidths that can be utilized in lasing\cite{zhou_lasing_2013,van_beijnum_surface_2013,meng_wavelength-tunable_2013,schokker_statistics_2015,dridi_lasing_2015,hakala_lasing_2017,RAM-17,REK-18} and Bose-Einstein condensation\cite{HakalaBEC2018}. Recently, it was demonstrated that collective SLR modes can be excited also in arrays of higher-loss ferromagnetic nanoparticles\cite{kataja_surface_2015,MAC-16}. 

Here, for the first time, we report on lasing in a lattice of \textit{ferromagnetic} nanodisks overlaid with optically pumped organic Rhodamine 6G (R6G) dye solution. Reduced linewidths provided by the SLRs together with a carefully optimized lattice geometry and gain medium produce lasing at visible wavelengths, despite the broad plasmonic resonances of the individual nanodisks. Lasing is characterized by a highly directional and nonlinear increase of sub-nanometer linewidth emission by more than two orders of magnitude. Within the limits set by the gain profile of R6G, the lasing wavelength can be tuned by varying the particle periodicity. In rectangular arrays, we observe lasing at two wavelengths corresponding to ${\lambda}\approx{n}{\times}p_{i}$ for different particle periodicities ($p_{x}$ and $p_{y}$) along the two principle axes of the lattice. 

In this study, we used Ni as ferromagnetic plasmonic material. We fabricated various arrays of Ni nanodisks on glass substrates using electron-beam evaporation and lift-off in an electron-beam lithography process. The nominal diameter and height of the nanodisks were \SI{60}{\nano\metre} and the total arrays size was \SI{300 x 300}{\micro\metre}. As reference, we considered a square array with $p_{x}=p_{y}=\SI{380}{\nano\metre}$. In rectangular arrays with broken symmetry, $p_{x}$ was kept constant and $p_{y}$ was varied from \SI{370}{nm} to \SI{390}{nm} in \SI{5}{nm} steps. The nanodisk arrays were covered by \SI{2}{nm} of Al$_2$O$_3$ using atomic layer deposition to protect Ni from degrading when contacted by R6G molecules. The gain medium consisting of 35 mM R6G in dimethyl sulfoxide (DMSO):benzyl alcohol (BA) (1:2) was inserted between the substrate with Ni nanodisk arrays and a cover glass.

\begin{figure}
\includegraphics[width=9cm]{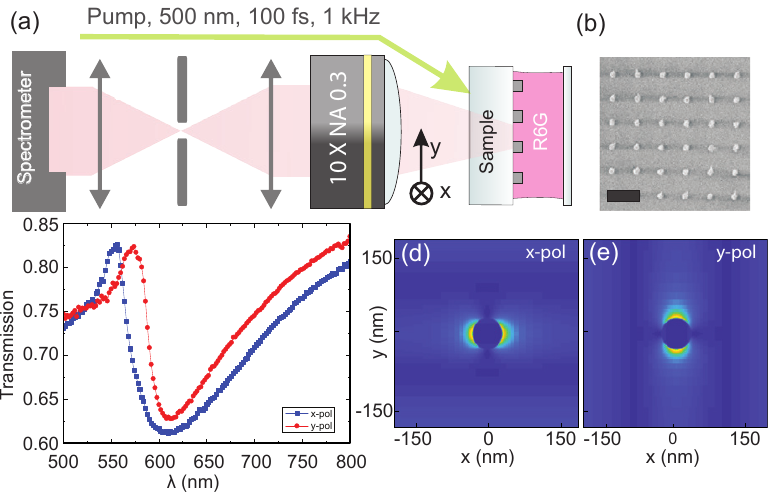}
\caption{\label{fig:Fig1} (a) Schematic of the measurement setup. Emission spectra are measured by focusing the back focal plane of the objective lens to the entrance slit of the spectrometer. The gain medium is pumped by $x$-polarized \SI{100}{fs} laser pulses with a wavelength of \SI{500}{nm} and a \SI{1}{kHz} repetition rate. The gain medium is inserted between the substrate with Ni nanodisk arrays and a cover glass. (b) Scanning electron microscopy (SEM) image of the \SI{380 x 380}{nm} array. In all experiments, the diameter and height of the Ni nanodisks are \SI{60}{nm}. The scale bar corresponds to \SI{500}{nm}. (c) Experimental transmission curves for the Ni nanodisk array with $p_{x}=\SI{380}{nm}$ and $p_{y}=\SI{370}{nm}$. Data for incident polarization along the $x$ and $y$ directions of the array are shown. (d),(e) Finite-difference time-domain (FDTD) simulations of near-field distributions in the same array. The simulations are performed for $x$- and $y$-polarized plane-wave excitation at the SLR wavelength.}
\end{figure}

A schematic of the measurement setup including excitation and detection lines is depicted in Figure~\ref{fig:Fig1}(a). The sample was excited by 100 fs laser pulses with a wavelength of \SI{500}{nm} at a \SI{1}{kHz} repetition rate and from a 45$^\circ$ angle. Emitted light from the sample was collected with a $10\times$ 0.3 NA objective. The back focal plane of the objective was focused to the entrance slit of a spectrometer. The long axis of the slit was aligned along the $y$-axis of the sample. From 2D intensity data collected by the CCD camera of the spectrometer, the wavelength and in-plane $k_y$-vector were calculated using $k_y=k_0\sin(\theta)$\cite{hakala_lasing_2017}. Here $k_0=2\pi/\lambda$ and $\theta$ is the angle with respect to the sample normal. We note that we did not apply a magnetic field during the reported lasing experiments. A scanning electron microscopy (SEM) image of the square \SI{380 x 380}{nm} reference array is shown in Figure~\ref{fig:Fig1}(b).   

Figure \ref{fig:Fig1}(c) shows experimental transmission curves of the Ni nanodisk array with $p_{x}=\SI{380}{nm}$ and $p_{y}=\SI{370}{nm}$ for incident polarization along $x$ and $y$ (see Supporting Information, Figure S1 for measurements on other arrays). The intensity maxima correspond to the DOs of the lattice. Since the DO wavelength depends on the particle periodicity perpendicular to the polarization axis, the transmission curve for $x$-polarized light is blue-shifted with respect to the spectrum measured with $y$ polarization. Coupling of a narrow DO to a broad localized surface plasmon resonance (LSPR) in the Ni nanodisks produces a collective SLR mode\cite{kataja_surface_2015}. The SLR wavelength (minimum transmission in Figure~\ref{fig:Fig1}(c)) corresponds to the wavelength where $1/\alpha-S$ is zero\cite{auguie_collective_2008,humphrey_plasmonic_2014}. Here, $S$ is the so-called array factor and $\alpha$ is the polarizability of a single Ni nanodisk. Damping of the SLR mode depends sensitively on the imaginary part of $1/\alpha$. Consequently, noble metal nanodisks with large polarizability produce narrow SLR modes with linewidth < 10 nm when ordered into periodic arrays\cite{zou_silver_2005,KRA-08,auguie_collective_2008,humphrey_plasmonic_2014,WANG-18,KRA-18}. Because of larger ohmic losses in Ni (i.e. small $\alpha$), the SLRs in our plasmonic arrays are much broader (> 100 nm, see Figure~\ref{fig:Fig1}(c)). Finite-difference time-domain (FDTD) simulations of the \SI{380 x 370}{nm} Ni nanodisk array at the SLR wavelengths (Figures~\ref{fig:Fig1}(d),(e)) show intense electric near fields at the particle plane, confirming the plasmonic character of this collective mode. Next, we demonstrate that despite the large linewidth of SLR excitations, it is possible to realize lasing in ferromagnetic nanostructures.

\begin{figure}
\includegraphics[width=9cm]{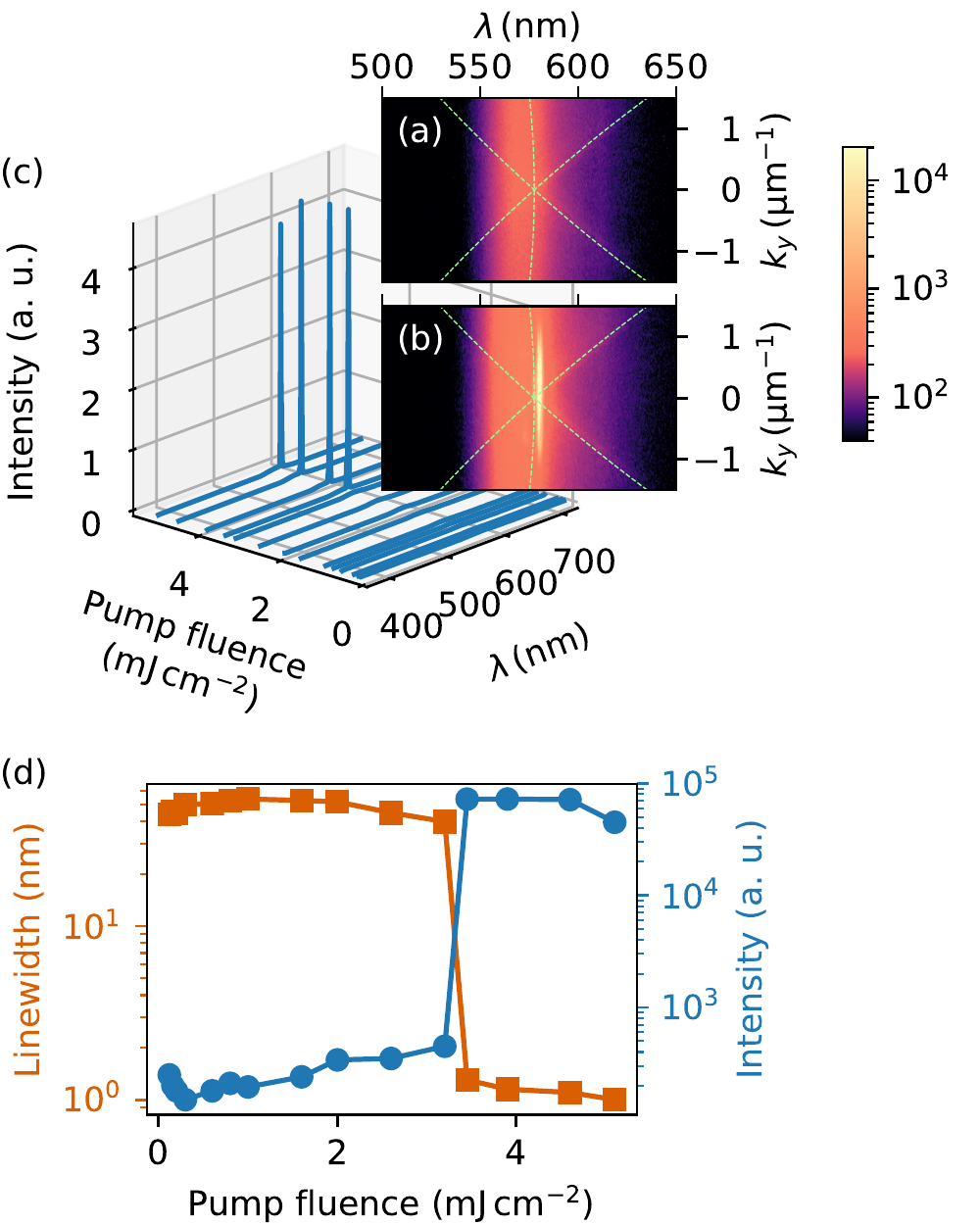}
\caption{\label{fig:Fig2} Angle and wavelength resolved emission of a symmetric \SI{380 x 380}{nm} Ni nanodisk array below (a) and above (b) the lasing threshold. The dashed lines indicate the DOs of the array. Since the momentum direction $k_y$ is monitored here, the $\left\langle+1,0\right\rangle$ and $\left\langle-1,0\right\rangle$ DOs related to the periodicity $p_y$ appear as a cross feature, while the one related to $p_x$ (around 580 nm) has a parabolic shape. The former is sometimes called TE and the latter TM mode in the literature\cite{BOZ-17}. (c) Emission intensity at $k_\mathrm{y}=0$ as a function of pump fluence. (d) Linewidth (squares) and intensity (circles) of the emission peak showing an abrupt nonlinear change of these parameters at a threshold pump fluence $P_\mathrm{th}\approx{3.3}$ mJ cm$^{-2}$.}
\end{figure}

Figures~\ref{fig:Fig2}(a),(b) show the $k_y$ and wavelength resolved emission from the sample with a $\SI{380 x 380}{nm}$ Ni nanodisk array for a pump fluence below and above the lasing threshold ($P_\mathrm{th}\approx{3.3}$ mJ cm$^{-2}$). Below threshold, the emission consists of two contributions. The first contribution has no angular dependence and originates from the R6G molecules. Because the molecules are spatially far away from the nanoparticles (and outside the SLR mode volume), they do not emit to the SLR mode. Their emission spectrum has a linewidth of $\sim$ 60 nm, which is the same as for the R6G dye solution in the absence of the nanodisk array. The second emission contribution follows the $\left\langle+1,0\right\rangle$ and $\left\langle-1,0\right\rangle$ DOs of the array (crossed dashed lines) and, thus, depicts spontaneous emission of the molecules to the SLR modes. As expected from ${\lambda}=n{\times}p_{i}$ and the transmission curves of Figure~\ref{fig:Fig1}(c), the DOs related to the periodicity $p_y$ cross at $\lambda=\SI{578}{nm}$ if $p_{y}=\SI{380}{nm}$ and $n=1.52$. 
At a higher pump fluence of $1.3P_\mathrm{th}$, we observe an intense single emission peak at $\lambda\approx\SI{580}{nm}$ with a narrow linewidth $<\SI{1}{nm}$ and a small beam divergence of \ang{5.7}. The lasing peak is slightly red shifted from the DOs to a wavelength where the R6G dye solution can emit to the SLR modes of the Ni nanodisk array. The transition from spontaneous emission to lasing is manifested as an abrupt change in the emission spectrum (Figure~\ref{fig:Fig2}(c)). Figure~\ref{fig:Fig2}(d) summarizes the variation of the emission intensity and linewidth with increasing pump fluence. Most notably, we measure a strongly nonlinear increase of the emission intensity from $\sim 2 \times 10^2$ to $\sim 10^5$ if the pump fluence is enhanced from 3.2 mJ cm$^{-2}$ to 3.45 mJ cm$^{-2}$. Simultaneously, the linewidth of the emission peak drops from $\sim \SI{60}{nm}$ to $<\SI{1}{nm}$.  

\begin{figure}
\includegraphics[width=9cm]{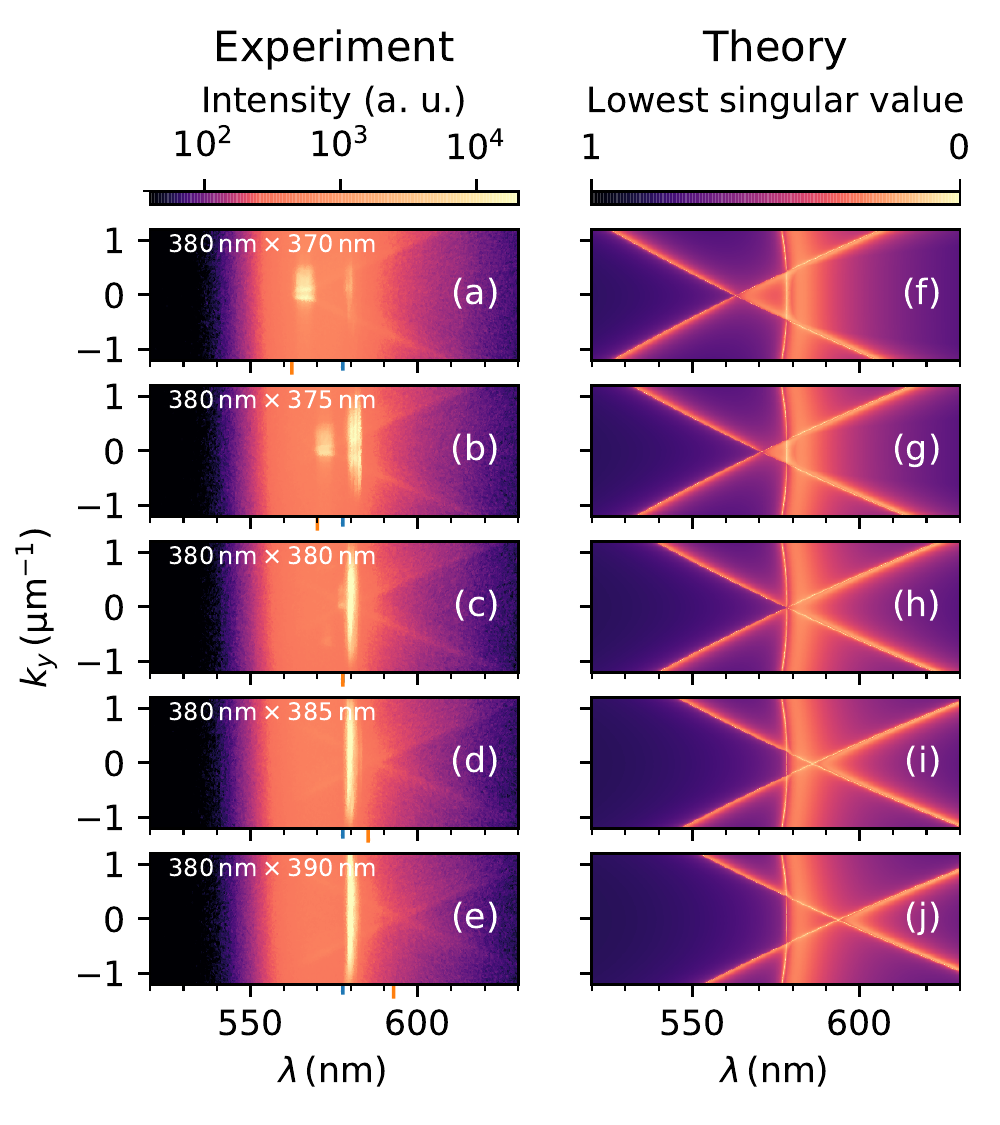}
\caption{\label{fig:Fig3}
(a–e) Angle and wavelength resolved emission data for samples having $p_x=380$ nm and $p_y$ ranging from \SI{370}{nm} to \SI{390}{nm} at a pump fluence of 4.6 mJ cm$^{-2}$. The red ticks label the crossing wavelength of the $\left\langle-1,0\right\rangle$ and $\left\langle+1,0\right\rangle$ diffracted orders related to the periodicity $p_y$ of the array. (f–j) Dispersions of respective ideal infinite arrays computed using the $T$-matrix method for $\mathbf{E}$-in-plane modes. Dispersion bands of the array are characterized by singular values of the underlying scattering problem (\ref{eq:SV problem}) reaching near-zero.}
\end{figure}

Next, we break the symmetry of the Ni nanodisk array by keeping $p_{x}$ constant and increasing $p_{y}$ from \SI{370}{nm} to \SI{390}{nm} in \SI{5}{nm} steps. Figure~\ref{fig:Fig3} shows emission spectra of these samples for a pump fluence of 4.6 mJ cm$^{-2}$. In (a), we observe two emission maxima, one at the same wavelength $\lambda\approx\SI{580}{nm}$ as for the square array (see Figure~\ref{fig:Fig3}(c)), and, the other at ${\lambda}\approx\SI{565}{nm}$. Threshold behavior and linewidth imply lasing action for both peaks. We note that the maximum at \SI{565}{nm} is \SI{15}{nm} blue-shifted from the other lasing peak. We associate this emission with the reduced particle periodicity along the $y$ direction. The expected 15 nm blue-shift based on $\Delta\lambda={n}\times\Delta{p_y}$ supports this argument, as well as the FDTD simulations of Supporting Information, Figure S3. The emission maximum at \SI{580}{nm} associates with the larger periodicity of the rectangular Ni nanodisk array along the $x$ direction. 

In agreement with the dependence of the two lasing peaks on particle periodicity, the lower-wavelength emission maximum red shifts when $p_y$ increases to \SI{375}{nm}, while the other peak remains fixed at \SI{580}{nm} (Figure~\ref{fig:Fig3}(b)). A further increase of $p_y$ to \SI{380}{nm} results in a square nanodisk lattice and, consequently, the two emissions merge into one intense lasing peak (Figure~\ref{fig:Fig3}(c)). For $p_y>$ \SI{380}{nm}, one would expect a second lasing peak to appear at $\lambda>$ \SI{580}{nm}. As can be seen from the emission data in Figure~\ref{fig:Fig3}(d),(e), this is not the case. We explain the absence of the anticipated second peak by a reduced overlap of the SLR mode with the wavelength-dependent gain profile of the R6G dye solution. At wavelengths corresponding to $p_y>$ \SI{380}{nm}, the R6G gain is insufficient to compensate for the lossy SLR mode of the Ni nanodisk array, resulting in much weaker spontaneous emission instead of lasing. For $p_y<$ \SI{380}{nm}, the spectrum of the R6G gain medium overlaps more with the energy of the SLR mode along the $y$ direction and, consequently, multimode lasing is observed. Previous studies on nanoparticle lattices made of noble metals include square arrays exhibiting lasing in both bright (dipolar) and dark (quadrupolar) modes, as well as broken symmetry superlattices in which lasing modes depend on the polarization of the pump pulse\cite{hakala_lasing_2017,Wang2017}. 

To understand the mode properties we employ a multiple scattering $T$-matrix approach\cite{guoLasingPointsHoneycomb2019}, which gives a characterization of the lattice modes in terms of a linear problem 
\begin{equation}
        M(\omega, \vect k) a_\nu(\omega, \vect k) = 0,
        \label{eq:SV problem}
\end{equation} where $M(\omega, \vect k)$
is a matrix depending on scattering and wave propagation properties of the array at a given frequency $\omega$
and $a_\nu(\omega, \vect k)$ is a vector of coefficients describing multipole nanoparticle excitations of a given mode. The problem (\ref{eq:SV problem}) has a nontrivial
solution, i.e., a mode exists, if the matrix $M(\omega, \vect k)$ has a zero singular value (SV). Therefore, to find modes supported by the array, we scan $\omega, \vect k$ space to search for SV minima of $M(\omega, \vect k)$. Due to losses, SV minima are not exactly zero for real $\omega, \vect k$, but near-zero SVs nevertheless provide valuable information about the dispersion of the array. Figures~\ref{fig:Fig3}(f)–(j) illustrate calculated modes with the lowest SVs of $M(\omega, \vect k)$, resembling the experimental data of Figures~\ref{fig:Fig3}(a)–(e). See Supporting Information for more details about the numerical model.

\begin{figure}
\includegraphics[width=9cm]{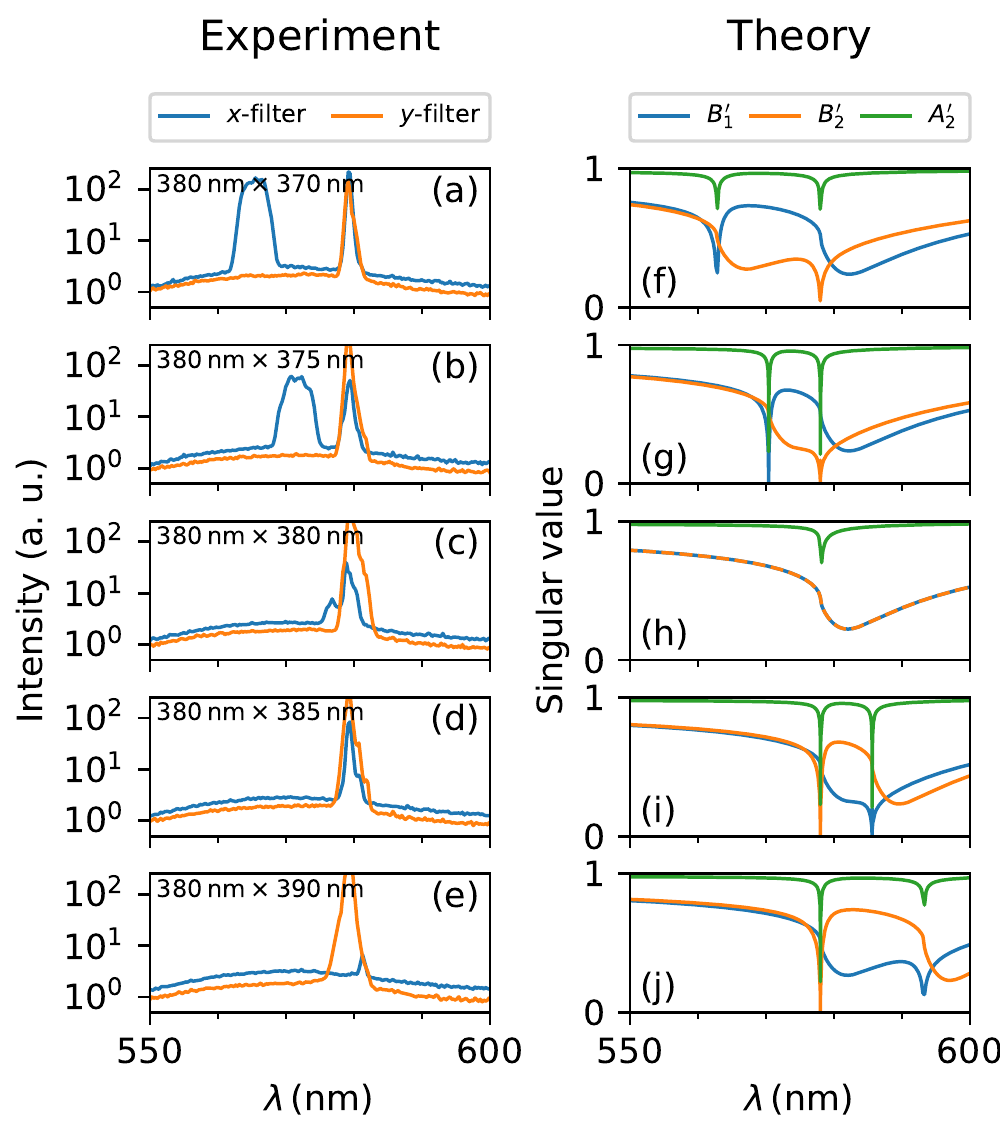}
\caption{\label{fig:Fig4} (a-e) Emission spectra at $k_y=0$ for samples having $p_x=\SI{380}{nm}$ and $p_y$ ranging from \SI{370}{nm} to \SI{390}{nm}. The pump fluence is 4.6 mJ cm$^{-2}$. (f-j) Calculated singular values for arrays with the same particle periodicities. The blue and orange colors correspond to $x$- and $y$-polarized dipolar modes ($B'_{1,2}$), respectively, and the green color corresponds to a quadrupolar mode ($A'_2$).}
\end{figure}

Finally, we analyze the polarization properties of the lasing modes in our Ni nanodisk samples with a R6G gain medium. Experimentally this was done by placing a polarizer between the sample and spectrometer to determine the emission intensity with polarization along the $x$ and $y$ directions of the arrays. Figures \ref{fig:Fig4}(a)–(e) present emission spectra for both polarization states at $k_y=0$ and a pump fluence of 4.5 mJ cm$^{-2}$. In (a), the low wavelength emission related to $p_y$ is $x$-polarized whereas the $p_x$-related lasing peak at \SI{580}{nm} exhibits both polarizations. We note that the $x$-polarization of this peak disappears at higher pump fluence (see Supporting Information, Figures S4 and S5). We observe similar behavior in (b), with a red shift in the lower wavelength lasing peak because of larger $p_y$. For the square array (c), we observe lasing at a single wavelength due to mode degeneracy. In (d) lasing takes place at the wavelength of the $p_x$-related SLR mode only. The lasing peak in this emission spectrum exhibits both polarizations. Finally, in (e) single mode lasing with reduced $x$ polarization is measured. 

The polarization properties are also studied by the the $T$-matrix method, which can uncover both $x$- and $y$-polarized dipolar as well as quadrupolar contributions at each wavelength. Figures \ref{fig:Fig4}(f)–(j) summarize the results for the different Ni nanodisk arrays. In (f) SV minima are calculated for two dipolar modes corresponding to the experimentally observed wavelengths in (a) and polarizations at high pump fluence. The same applies for (g). In (h), the square array exhibits degeneracy of the dipolar modes. This raises the question whether the experimental observations in (c) are caused by lasing action of two perpendicular dipolar modes or a quadrupolar mode. Finally, in (i) and (j), we note that the model predicts a large quadrupolar weight at the measured lasing wavelength in (d) and (e). In (d), the experimental lasing peak exhibits both polarizations and, hence, the mode is indeed quadrupolar. In contrast, lasing in (e) is almost purely $y$-polarized, suggesting dipolar mode lasing. To rationalize this, we point out that a priori predictions of preferred lasing modes are difficult because mode dynamics, mode competition at available gain, and mode Q-factors all play a role. Furthermore, we observe a dependence of lasing behavior on pump fluence (see Supporting Information, Figure S4 for a complete set of emission spectra at different pump fluence). A more detailed understanding of mode competition in ferromagnetic nanodisk arrays requires further studies. Here, our main result is the first demonstration of lasing in a high-loss ferromagnetic plasmonic system. In rectangular Ni nanodisk arrays, we observe multimode lasing and the coexistence of dipolar and quadrupolar modes. 

The results of this paper pave the way for incorporating gain into novel magnetoplasmonic devices and realizing new concepts for topological photonics. Notably, topological lasing has been demonstrated recently\cite{BAH-17,HAR-18,BAN-18}. In topological photonics\cite{KHA-17,Haldane2008,Lu2016,Klembt2018}, most lattice systems are based on nearest or next-nearest neighbor coupling via overlapping optical near fields. Our ferromagnetic nanodisk arrays represent a radiatively coupled system where long-range couplings produce collective SLR modes. The symmetry properties of the array dictate the existence of energy degenerate modes at high-symmetry points of the Brillouin zone\cite{guoLasingPointsHoneycomb2019}, for which the lifting of the degeneracy by a symmetry breaking mechanism can lead to topological features. The magnetic moment of nanodisks in a ferromagnetic array can be exploited as a new tool for time-reversal symmetry breaking in such lattices.      

\section{Associated Content}
\subsection{Supporting Information}
Transmission curves for all Ni nanodisk arrays. Details on FDTD lasing simulations and mode calculations using the multiple scattering $T$-matrix approach. Dependence of emission spectra on the laser pump fluence including polarization analysis of the lasing peaks.

\subsection{Notes}
The authors declare no competing financial interest.

\begin{acknowledgement}
This work was supported by the Academy of Finland under Projects No. 303351, No. 307419, No. 318987, No. 316857, by the European Research Council (ERC-2013-AdG-340748-CODE), and by the Aalto Centre for Quantum Engineering. Lithography was performed at the Micronova Nanofabrication Centre, supported by Aalto University.
\end{acknowledgement}


\providecommand{\latin}[1]{#1}
\makeatletter
\providecommand{\doi}
{\begingroup\let\do\@makeother\dospecials
	\catcode`\{=1 \catcode`\}=2 \doi@aux}
\providecommand{\doi@aux}[1]{\endgroup\texttt{#1}}
\makeatother
\providecommand*\mcitethebibliography{\thebibliography}
\csname @ifundefined\endcsname{endmcitethebibliography}
{\let\endmcitethebibliography\endthebibliography}{}

\end{document}